\documentstyle[12pt]{article}
\begin{document}

\title{Small oscillations of a Chiral Gross-Neveu system.}

\author{P.L. Natti\thanks{Supported by Funda{\c{c}\~ao} de Amparo \`a 
Pesquisa do Estado de S\~ao Paulo (FAPESP), Brazil} 
and A.F.R. de Toledo Piza \\
Instituto de F{\'{\i}}sica, Universidade de S\~ao Paulo,\\
C.P. 66318, CEP 05389-970 S\~ao Paulo, S.P., Brasil}

\maketitle

\begin{abstract}

We study the small oscillations regime (RPA approximation) of the
time-dependent mean-field equations, obtained in a previous work
\cite{NTP}, which describe the time evolution of one-body dynamical
variables of a uniform Chiral Gross-Neveu system. In this
approximation we obtain an analytical solution for the time evolution
of the one-body dynamical variables. The two-fermion physics can be
explored through this solution. The condition for the existence of
bound states is examined.

\vskip 1.0cm

PACS Numbers : 03.65.Ge, 03.65.Nk, 11.30Rd, 11.10Kk

\end{abstract}

\section{Introduction}
\vskip 0.7cm

In a previous work \cite {NTP} we obtained in Gaussian mean-field
approximation the effective dynamics of one-fermion and pairing
densities of a off-equilibrium spatially uniform (1+1) dimensional
self-interacting fermion system described by Chiral Gross-Neveu model
(CGNM) \cite {GN}.  These dynamical equations acquire the structure of
(colisionless) kinetic equations.  They determine the time evolution
of the all one-fermion densities of this system for a given initial
condition. Spatial uniformity (translation and reflection invariance)
is assumed in our derivation.

Studing the static solutions of these equations in order to
renormalize the theory \cite {NTP}, we found an effective potential
similar to that obtained by Gross and Neveu using the $1/N$ expansion
\cite {GN}. We also showed that other static results which have been
discussed in the literature \cite {GN,DHN,RK} such as dynamical mass
generation due to chiral symmetry breaking and a phenomenon analogous
to dimensional transmutation can be retrieved from this
formulation. Finally, in \cite {NTP}, we obtained numerical solutions
for the time evolution of the one-body dynamical variables initially
displaced from equilibrium. The time evolution of the symmetrical and
broken chiral phases of our system are discussed.

In this work we explore a particular application of the renormalized
nonlinear mean-field equations obtained in \cite {NTP}. We follow the
recent work of Kerman and Lin \cite{KL} in order to study the near
equilibrium dynamics around the stationary solution as a tool to
investigate the two-fermion dynamics. In particular, the resulting
equations can be solved analytically to reveal a two-(quasi)fermion
bound state solution.

This paper is organized as follows. In Sec.II we linearize the mean-field 
dynamical equations which describe the time evolution of a off-equilibrium 
spatially uniform (1+1) dimensional self-interacting fermion system described by Chiral Gross-Neveu model (CGNM). 
A self-consistent renormalization scheme 
is necessary \cite {NTP,TOM}. In Sec.III, making use of an 
analogy with scattering theory \cite {RN}, we 
obtain a closed analytical solution for the time evolution of 
the one-fermion densities in this regime. 
Studyng the two-fermion physics in Sec.IV, we find the condition for the 
existence of bound states. Finally, the Sec.V is devoted to a final 
discussion and conclusions.

\vskip 1.0cm
\section{Mean-field kinetic equations}
\vskip 0.7cm

We begin this section by reviewing  
our approach which describe a formal 
treatment of the kinetics of a self-interacting 
quantum field. This approach 
was developed earlier for the nonrelativistic 
nuclear many-body dynamics by 
Nemes and de Toledo Piza \cite {Nemes} and was more 
recently applied in the 
quantum-field theoretical context to the 
self-interacting $\lambda \phi^{4}$ 
theory in (1+1) dimensions \cite{Lin}. The general  
ideia is to focus on the time evolution 
of the one-fermion and pairing densities. These 
observables are kept under direct control 
when one works variationally 
using a Gaussian functional {\it ansatz} and 
will therefore be refered to 
as Gaussian observables.

We consider an off-equilibrium, spatially uniform, 
(1+1) dimensional system of 
relativistic, self-interacting fermions 
described by Chiral Gross-Neveu model 
(CGNM) \cite {GN}. The Hamiltonian density is given by

\begin{equation}
\label{1}
{\cal H}_{\mbox{\tiny CGNM}}=\sum_{i=1}^{N}\left\{\bar\psi^{i}
\left[-i\gamma_{1}\partial_{1}\right]\psi^{i}\right\}-
\frac {g^{2}}{2}\left\{\left[\sum_{i=1}^{N}
\bar\psi^{i}\psi^{i}\right]^{2}-
\xi\left[\sum_{i=1}^{N}\bar\psi^{i}
\gamma_{5}\psi^{i}\right]^{2}\right\}\;\;,
\end{equation}
\vskip 0.7cm

\noindent
where $\xi$ is a constant which indicates whether 
the model is invariant under 
discrete $\gamma_{5}$ transformation ($\xi=0$) or under the 
Abelian chiral $U(1)$ group ($\xi=1$). 
In the form considered here, this is a 
massless fermion theory in (1+1) 
dimensions with quartic interaction. 
The model contains $N$ species of 
fermions coupled symmetrically. In the 
Heisenberg picture, the $\psi^{i}$ are complex Dirac spinors 

\begin{eqnarray}
\label{2}
\psi(x)&=&\sum_{\bf k}\left(\frac {m}{k_{0}}\right)^{1/2}
\left[b_{{\bf k},1}(t)u_{1}({\bf k})\frac 
{e^{i{\bf kx}}}{\sqrt{L}}+b^{\dag}_{{\bf k},2}
(t)u_{2}({\bf k})\frac {e^{-i{\bf kx}}}{\sqrt{L}}\right]
\nonumber\\ \\
\bar\psi(x)&=&\sum_{\bf k}\left(\frac{m}{k_{0}}\right)^{1/2}
\left[b^{\dag}_{{\bf k},1}(t)\bar u_{1}({\bf k})
\frac {e^{-i{\bf kx}}}{\sqrt {L}}+b_{{\bf k},2}(t)\bar u_{2}({\bf k})
\frac {e^{i{\bf kx}}}{\sqrt {L}}\right] \;\;,
\nonumber \\ \nonumber
\end{eqnarray}

\noindent
where $b^{\dag}_{{\bf k},1}$ and $b_{{\bf k},1}$ [$b^{\dag}_{{\bf k},2}$ and
$b_{{\bf k},2}$] are fermion creation and annihilation operators
associated to positive [negative]-energy solution $u_{1}({\bf k})$
[$u_{2}({\bf k})$] of Dirac's equation.

This model is essentially equivalent to the Nambu-Jona-Lasinio model
\cite{NJ}, except for the fact that in (1+1) dimensions it is
renormalizable. Moreover, it is one of the very few known field
theories which are assymptotically free. To leading order in $1/N$
expansion \cite{GN}, the CGNM exhibits a number of interesting
phenomena, like spontaneous symmetry breaking, dynamical fermion mass
generation and dimensional transmutation.

The state of this system (assumed spatially uniform) is given in terms
of a many-body density operator ${\cal F}$ of unit trace.  Our
implementation of the Gaussian mean-field approximation consists in
approximating this object by a truncated many-body density operator
${\cal F}_{0}(t)$, also of unit trace, written as the most general
hermitian Gaussian functional of the field operators consistent with
the assumed uniformity of the system \cite{DES}. It will thus be
written as the exponential of a general quadratic form in the field
operators, which can be reduced to diagonal form by suitable canonical
transformation. The most general transformation would in general break
both the chiral and charge symmetries of the CGNM. In the following
development we restrict ourselves for simplicity to a special class of
transformations (to be called Nambu transformations) which break the
chiral symmetry only. They can be parametrized in a form that
incorporates the unitarity constraints as

\begin{equation}
\label{3}
\left[
\begin{array}{c}
      b_{{\bf k},1}    \\
                   \\
      b_{{\bf k},2}    \\
                   \\
b_{-{\bf k},1}^{\dag}\\
                   \\
b_{-{\bf k},2}^{\dag}
\end{array}
\right]=\left[
\begin{array}{cccc}
\cos\varphi_{\bf{k}} &      0       &      0       & 
          -e^{-i\gamma_{\bf{k}}}\sin\varphi_{\bf{k}} \\
          &              &              &              \\  
  0       & \cos\varphi_{\bf{k}} & e^{-i\gamma_{\bf{k}}}
\sin\varphi_{\bf{k}} &     0        \\ 
          &              &              &              \\ 
  0       & -e^{i\gamma_{\bf{k}}}\sin\varphi_{\bf{k}}
          &\cos\varphi_{\bf{k}} &     0        \\ 
          &              &              &              \\
e^{i\gamma_{\bf{k}}}\sin\varphi_{\bf{k}} &      0       &      0
          & \cos\varphi_{\bf{k}} 
\end{array}
\right]\left[
\begin{array}{c}
    \beta_{{\bf k},1}    \\
                         \\ 
    \beta_{{\bf k},2}    \\
                         \\
\beta_{-{\bf k},1}^{\dag}\\
                         \\
\beta_{-{\bf k},2}^{\dag}
\end{array}
\right]
\end{equation}

\vskip 0.5cm
\noindent
where reflection symmetry of the uniform system is implemented by
making the parameters $\varphi_{\bf{k}}$ and $\gamma_{\bf{k}}$
dependent only on the magnitude of ${\bf k}$.

The Gaussian truncated density operator ${\cal F}_{0}(t)$ acquires a
particularly simple form when expressed in terms of the Nambu
quasi-fermion operators which diagonalize the associated quadratic
form, namely

\begin{equation}
\label{6}
{\cal F}_{0}(t)=\prod_{{\bf k},\lambda}
\left[ \nu_{{\bf k},\lambda}
\beta_{{\bf k},\lambda}^{\dag}(t)
\beta_{{\bf k},\lambda}(t)+
(1-\nu_{{\bf k},\lambda})
\beta_{{\bf k},\lambda}(t)
\beta_{{\bf k},\lambda}^{\dag}(t)\right]\;\;,
\end{equation}

\vskip 0.5cm
\noindent where $\nu_{{\bf k},\lambda}$ for $\lambda=1,2$ 
are the Nambu (quasi-fermion) occupation numbers.

With the help of Eq.(\ref{3}) it is an easy task to express
$\bar\psi(x)$ and $\psi(x)$, Eq.(\ref{2}), in terms of $\beta_{{\bf
k},\lambda}^{\dag}(t)$ and $\beta_{{\bf k},\lambda}(t)$ for $
\lambda=1,2$. In doing so, one finds that the plane waves of
$\bar\psi(x)$ and $\psi(x)$ are modified by a complex,
moment-dependent redefinition of $m$ involving the Nambu parameters
$\varphi_{\bf k}(t)$ and $\gamma_{\bf k}(t)$. The complex character of
these parameters is actually crucial in dynamical situations, where
the imaginary parts will allow for the description of time-odd
(velocity-like) properties. Finally, the mean values of the Gaussian
observables are parametrized in terms of the $\varphi_{\bf k}(t)$ and
$\gamma_{\bf k}(t)$ and of the occupation numbers $\nu_{{\bf
k},\lambda}(t)= Tr \left[\beta^{\dag}_{{\bf k},\lambda}(t) \beta_{{\bf
k},\lambda}(t){\cal F}(t)\right]$ for $ \lambda=1,2$.

The next step is to obtain the mean-field time evolution for the mean
values of the Gaussian observables in the context of the initial-value
problem.  In other words, we want the Gaussian mean-field equations of
motion for the Nambu parameters $\varphi_{\bf k}(t)$ , $\gamma_{\bf
k}(t)$ and for the quasi-particle occupation numbers $\nu_{{\bf
k},\lambda}(t)$. In Ref.\cite{NTP} we obtained

\begin{equation}
\label{7}
\dot\nu_{{\bf k},1}=0\;\;\;{\rm and}\;\;\;\dot\nu_{{\bf k},2}=0\;\;\;;
\end{equation}

\begin{equation}
\label{8}
\left[i\dot\varphi_{\bf k}+\dot\gamma_{\bf k}\sin\varphi_{\bf k}
\cos\varphi_{\bf k}\right]e^{-i\gamma_{\bf k}}=
\frac {Tr\left([\beta_{-{\bf k},1}\beta_{{\bf k},2},
H_{\mbox{\tiny CGNM}}]{\cal F}_{0}\right)}
{(1-\nu_{{\bf k},1}-\nu_{{\bf k},2})}\;\;.
\end{equation}
\vskip 0.5cm

Eq.(\ref{7}) shows that the occupation numbers of the Nambu
quasi-particles are constant. This is the general isoentropic
character of the mean-field approximation \cite{Carol,Eboli}. The
complex equation of motion (\ref{8}) describes the time evolution of
the Nambu parameters. From the right-hand side of the Eq.(\ref{8}), we
see that to obtain the time evolution of the Nambu parameters, we have
to express the CGNM Hamiltonian in the Nambu basis.

From the Hamiltonian density (\ref{1}) we can explicitly evaluate the
Hamiltonian of the system by integration over all one-dimensional
space. This involves, in particular, choosing a representation for the
$\gamma$-matrices. Here we have to be careful, since a bad choice of
representation can spoil manifest reflection invariance (see Appendix
A of Ref.\cite {NTP}).  We choose the Pauli-Dirac representation,
namely

\begin{equation}
\label{9}
\gamma_{0}=\sigma_{3}\;\;\;;\;\;\;
\gamma_{1}=i\sigma_{2}\;\;\;{\rm and}\;\;\;
\gamma_{5}=\gamma_{0}\gamma_{1}=\sigma_{1}\;\;.
\end{equation}
\vskip 0.5cm 
\noindent

Substituting the CGNM Hamiltonian written in Nambu basis in the
dynamical equation (\ref{8}), we obtain an explicit dynamical equation
which describes the time evolution of the Nambu parameters.  The
calculation of traces is lengthy but straightforward. Taking the case
$N=1$ for simplicity and splitting the resultant complex equation into
real and imaginary parts we have

\begin{eqnarray}
\label{10}
\dot\nu_{{\bf k},1}&=&0\;\;\;{\rm and}
\;\;\;\dot\nu_{{\bf k},2}=0\;\;\;;
\nonumber\\ \nonumber\\
\dot\varphi_{\bf k}&=&\sin\gamma_{\bf k}
\frac {|{\bf k}|}{k_{0}}\left[m-
\left(\frac {g^{2}m}{4\pi}\right)
(\xi+1)(I_{1}+I_{2})\right]\nonumber\\ \\
\dot\gamma_{\bf k}\sin 2\varphi_{\bf k}&=&\frac 
{2\sin 2\varphi_{\bf k}}{k_{0}}\left[{\bf k}^{2}+
\left(\frac {g^{2}m^{2}}{4\pi}\right)
(\xi+1)(I_{1}+I_{2})\right]+ 
\nonumber\\ \nonumber\\
&+&2\cos  2\varphi_{\bf k}
\cos\gamma_{\bf k}\frac {|{\bf k}|}{k_{0}}
\left[m-\left(\frac {g^{2}m}{4\pi}\right)
(\xi+1)(I_{1}+I_{2})\right]
\;\;,\nonumber\\ \nonumber
\end{eqnarray}
 
\noindent
where $I_{1}$ and $I_{2}$ are the divergent integrals

\begin{eqnarray}
\label{11}
I_{1}&=&\int \frac {d{\bf k}'}{k_{0}'}\cos 2\varphi_{{\bf k}'}
(1-\nu_{{\bf k}',1}-\nu_{{\bf k}',2})\nonumber\\ \\
I_{2}&=&\int \frac {d{\bf k}'}{k_{0}'}\frac {|{\bf k}'|}{m}
\sin 2\varphi_{{\bf k}'}\cos\gamma_{{\bf k}'}(1-\nu_{{\bf k}',1}-
\nu_{{\bf k}',2})\;\;.\nonumber\\ \nonumber
\end{eqnarray} 

A renormalization procedure is now required to deal with these
divergences.  In general, renormalization procedures consist in
combining divergent terms with the bare mass and coupling constants of
the theory to define new, finite (or renormalized) values of these
quantities.  In other words, the bare mass and coupling constants are
chosen to be cut-off dependent in a way that will cancel the
divergences. In the present case, however the divergent integrals
(\ref{11}) involve the dynamical variables themselves in the
integrand, so that even their degree of divergence is not directly
computable. In order to handle this situation we will use a
self-consistent renormalization procedure inspired in Ref.\cite{TOM}.

The renormalization prescription that we use is based on the
consideration of the static solutions of the dynamical equations
(\ref{10}), which satisfy

\begin{eqnarray}
\label{12}
&& \sin\gamma_{\bf k}|_{\mbox{\scriptsize eq}}
\left[1-\left(\frac {g^{2}}{4\pi}\right)
(\xi+1)(I_{1}+I_{2})\right]= 0 \\ \nonumber\\\nonumber\\
\label{13}
&& \tan 2\varphi_{\bf k}|_{\rm eq}=
\frac{-|{\bf k}|m\left[1-\left(g^{2}/4\pi\right)
(\xi+1)(I_{1}+I_{2})\right]}
{\left[({\bf k})^{2}+\left(g^{2}m^{2}/4\pi\right)
(\xi+1)(I_{1}+I_{2})\right]}
\cos\gamma_{\bf k}|_{\rm eq} \;\;.\\\nonumber
\end{eqnarray}
\vskip 0.5cm

\noindent
We will show explicitly that controlling the divergences of these
equations will also control the divergences which appear in the
kinetic regime of the mean-field approximation [see below,
Eqs.(\ref{19})].

In order to obtain the renormalization prescription we introduce a
regularizing momentum cut-off $\Lambda$ and begin by assuming that, in
order to render the theory finite, the bare coupling constant $g^2$
must approach zero for large values of the $\Lambda$ as (see
e.g. Refs.\cite{GN,RK})

\begin{equation}
\label{14}
g^{2}=\frac{4\pi}{(\xi+1)}
\left[\ln\left(\frac {\Lambda^{2}}{m^{2}}\right)\right]^{-1}\;\;,
\end{equation}
\vskip 0.5cm

\noindent
where the form of the first factor is dictated by later
convenience. We next {\it assume} that the integrals $I_{1}$ and
$I_{2}$ have logarithmic divergences

\begin{eqnarray}
\label{15}
I_{1}&=& a+b\ln\left(\frac {\Lambda^{2}}{m^{2}}\right)\nonumber\\ \\
I_{2}&=& c+d\ln\left(\frac {\Lambda^{2}}{m^{2}}\right)\nonumber\;\;,
\end{eqnarray}
\vskip 0.3cm

\noindent where $a$, $b$, $c$ and $d$ are finite constants.
Substituting (\ref{14}) and the {\it ansatze} (\ref{15}) in static
equations (\ref{12}) and (\ref{13}), we obtain

\vskip 0.2cm
\begin{eqnarray}
\label{16}
&& d\sin \gamma_{\bf k}|_{\rm eq}=0\nonumber\\\\
&& \tan 2\varphi_{\bf k}|_{\rm eq}=\frac {-(-1)^{n}m|{\bf k}|
[1-(b+d)]}{[({\bf k})^{2}+m^{2}(b+d)]}\nonumber\;\;.
\end{eqnarray}
\vskip 0.5cm

\noindent We now verify that the assumed divergent character of the
integrals $I_{1}$ and $I_{2}$ is consistent with the {\it ansatze}
(\ref{15}).  Substituting the solution (\ref{16}) into Eq.(\ref{11})
we find that $I_{1}$ and $I_{2}$ indeed have the prescribed
logarithmic divergence (see Appendix C of Ref.\cite {NTP}).  Moreover,
from this calculation, we obtain the values of the constants $a$, $b$,
$c$ and $d$.  We find $b=1$, while $d$ remains arbitrary.  The
renormalized static solution of our system in the mean-field
approximation are then obtained by simply substituting these values in
Eq.(\ref{16}):

\begin{eqnarray}
\label{17}
&&d\sin \gamma_{\bf k}|_{\rm eq}=0\nonumber\\\\
&&\tan 2\varphi_{\bf k}|_{\rm eq}=
\frac {(-1)^{n}m|{\bf k}|d}{[{\bf k}^{2}+
(1+d)m^{2}]}\;\;.\nonumber
\end{eqnarray}
\vskip 0.3cm

\noindent We also show in Ref.\cite{NTP} 
that the connection between particle mass $m$ and quasi-particle mass
$m_{\mbox{\scriptsize eff}}$ is given by

\begin{equation}
\label{18}
m_{\mbox{\scriptsize eff}}=(1+d)m\;\;.
\end{equation}
\vskip 0.5cm

\noindent From the redefinition of the mass scale given by 
Eq.(\ref{18}), we note that, unlike the situation found in connection
with the $1/N$ expansion, the use of the Gaussian {\it ansatz},
Eq.(\ref{6}), parametrized by the canonical transformation leading to
the quasi-fermion basis, allows for the direct dynamical
determination of the stable equilibrium situation of the system [see
Eq.(\ref{17})], including symmetry breaking [when $d\neq-1$, see also Ref.\cite{NTP}] and mass generation [see Eq.(\ref{18})]. 
Moreover, the renormalization procedure effectively
replaces the dimensionless coupling constant $g^2$ by the free
parameter $d$ associated to the mass scale [see Eq.(\ref{18})]. This
is analogous to the phenomenon of dimensional transmutation found by
Gross and Neveu \cite{GN} in the case of an $1/N$ expansion.  Finally,
aside from the over-all mass scale (characterized by $d$) there are no
free adjustable parameters.

Using Eqs.(\ref{10}), (\ref{14}) and (\ref{15}), we finally write the
renormalized form of the dynamical equations that describe the
mean-field time evolution of the system. As mentioned before they are
now also finite and read

\begin{eqnarray}
\label{19}
\dot\nu_{{\bf k},1}&=& 0\;\;\;{\rm e}\;\;\;
\dot\nu_{{\bf k},2}=0 
\nonumber\\ \nonumber\\
\dot\varphi_{\bf k}&=&(-1)md\frac {|{\bf k}|}{k_{0}}
\sin\gamma_{\bf k}
\nonumber\\ \\
\dot\gamma_{\bf k}\sin 2\varphi_{\bf k}&=&
\frac {2\sin 2\varphi_{\bf k}}{k_{0}}
\left[({\bf k})^{2}+m^{2}(1+d)\right]+
\nonumber\\ \nonumber\\
&-&2md\frac{|{\bf k}|}{k_{0}}\cos 2\varphi_{\bf k}
\cos\gamma_{\bf k}\;\;.
\nonumber 
\end{eqnarray}
\vskip 0.3cm

\subsection{Small oscillations regime}

In order to study the small oscillation regime of the kinetic
equations we next linearize the above kinetic equations around the
static solution (\ref{17}) taking $\nu_{{\bf k},1}=\nu_{{\bf k},2}=0$.
We begin by introducing the displacement away from equilibrium of the
dynamical variables $\varphi_{\bf k}$ and $\gamma_{\bf k}$

\begin{eqnarray}
\label{20}
\varphi_{\bf k}&=&\varphi_{\bf k}|_{\rm eq}+\delta\varphi_{\bf k}
\nonumber\\ \\
\gamma_{\bf k}&=&\gamma_{\bf k}|_{\rm eq}+\delta\gamma_{\bf k}\;\;,
\nonumber \\ \nonumber
\end{eqnarray}

\noindent
where the static solution $\varphi_{\bf k}|_{\rm eq}$ and 
$\gamma_{\bf k}|_{\rm eq}$ are obtained from the Eq.(\ref{17})
[see also Ref.\cite{NTP}]

\begin{eqnarray}
\label{21}
\sin 2\varphi_{\bf k}|_{\rm eq}&=&\frac{m|{\bf k}|d}{k_{0}
[{\bf k}^{2}+(1+d)^{2}m^{2}]^{1/2}} \nonumber \\ \nonumber\\
\cos 2\varphi_{\bf k}|_{\rm eq}&=&
\frac{[{\bf k}^{2}+(1+d)m^{2}]}{k_{0}
[{\bf k}^{2}+(1+d)^{2}m^{2}]^{1/2}} \\ \nonumber\\
\gamma_{\bf k}|_{\rm eq}&=&0\;\;\;\mbox{with}\;\;d\neq 0\;\;.
\nonumber \\ \nonumber
\end{eqnarray}

\noindent The quantities $\delta\varphi_{\bf k}$ and
$\delta\gamma_{\bf k}$ ,will be treated as (first-order) small
displacements.  Functions of the dynamical variables are expanded also
to first-order around the equilibrium solution (\ref{21}). Therefore,
we must linearize the divergent integrals (\ref{11}) around
equilibrium. Taking $\nu_{{\bf k},\lambda}=0$ we have

\begin{eqnarray}
\label{22}
I_{1}&=&I_{1}^{(0)}+I_{1}^{(1)}+{\cal O}(\delta\varphi_{\bf k})^{2}
\nonumber\\ \nonumber\\
&=&\int\frac{d{\bf k}'}{k_{0}'}\cos2\varphi_{{\bf k}'}|_{\rm eq}-
2\int\frac{d{\bf k}'}{k_{0}'}\sin 2\varphi_{{\bf k}'}|_{\rm eq}\;
\delta\varphi_{{\bf k}'}\nonumber\\\\ 
I_{2}&=&I_{2}^{(0)}+I_{2}^{(1)}+
{\cal O}[(\delta\varphi_{\bf k})^{2},(\delta\gamma_{\bf k})^{2},
(\delta\varphi_{\bf k}\delta\gamma_{\bf k})]\nonumber \\ \nonumber\\
&=&\int \frac{d{\bf k'}}{k_{0}'}\frac{|{\bf k}'|}{m}
\sin 2\varphi_{\bf k}|_{\rm eq}+2\int\frac{d{\bf k}'}{k_{0}'}
\frac{|{\bf k}'|}{m}\cos 2\varphi_{\bf k}|_{\rm eq}\;
\delta\varphi_{{\bf k}'} \;\;. \nonumber \\ \nonumber
\end{eqnarray}

The linearized form of kinetic equations for $\delta\varphi_{\bf k}$
and $\delta\gamma_{\bf k}$ are then obtained as

\begin{eqnarray}
\label{23}
&& \delta\dot\varphi_{\bf k}= - \frac{m|{\bf k}|d}{k_{0}}
\delta\gamma_{\bf k}\\ \nonumber\\
\label{24}
&& \delta\dot\gamma_{\bf k}\frac{m|{\bf k}|d}{k_{0}}=4[{\bf k}^{2}+
(1+d)^{2}m^{2}]\delta\varphi_{\bf k} + \nonumber\\ \\
&& - 4\left(\frac{g^{2}}{4\pi}\right)(\xi+1)|{\bf k}|\int d{\bf k}'
\frac{|{\bf k}'|}{[({\bf k}')^{2}+(1+d)^{2}m^{2}]^{1/2}}
\delta\varphi_{{\bf k}'}\;\;, \nonumber
\end{eqnarray}
\vskip 0.5cm

\noindent where the renormalization procedure (\ref{14}) 
controls the logarithmic divergence of the integral appearing in
(\ref{24}) [see below Eqs.(\ref{35}) and (\ref{39})].  Substituting
Eq.(\ref{24}) into Eq.(\ref{23}) we obtain finally

\begin{eqnarray} 
\label{25}
&& \delta\ddot\varphi_{\bf k}+4[{\bf k}^{2}+(1+d)^{2}m^{2}]
\delta\varphi_{\bf k} + \nonumber\\ \\
&& - 4\left(\frac{g^{2}}{4\pi}\right)(\xi+1)|{\bf k}|
\int d{\bf k}'\frac{|{\bf k}'|}{[({\bf k}')^{2}+(1+d)^{2}m^{2}]^{1/2}}
\delta\varphi_{{\bf k}'}=0. \nonumber
\end{eqnarray}
\vskip 0.5cm

\noindent As usual in small oscillation treatments, this is a linear
oscillator equation.  Note that the last term couples different
momenta.  The solution to this problem involves determining the normal
modes of small oscillation and their frequencies.  This is done by
looking for solutions of Eqs.(\ref{23}) and (\ref{24}), which are of
the form

\begin{eqnarray}
\label{26}
\delta\varphi_{\bf k}&=&\Psi_{\mbox{\scriptsize\bf k}}
\:e^{i\omega t}\nonumber \\ \\
\delta\gamma_{\bf k}&=&\Gamma_{\mbox{\scriptsize\bf k}}
\:e^{i\omega t}\;\;,\nonumber \\ \nonumber
\end{eqnarray} 

\noindent where $\Psi_{\mbox{\scriptsize\bf k}}$ 
and $\Gamma_{\mbox{\scriptsize\bf k}}$ are time-independent
amplitudes. Substituting the solution (\ref{26}) into Eqs.(\ref{23})
and (\ref{24}), we obtain equations for these amplitudes, namely

\begin{eqnarray}
\label{27}
&& i\omega\Psi_{\bf k}+\frac{m|{\bf k}|d}{k_{0}}\Gamma_{\bf k}=0 
\\ \nonumber\\
\;\;\;\;\;\;\;\;
\label{28}
&& i\omega\frac{m|{\bf k}|d}{k_{0}}\Gamma_{\bf k}-
4[{\bf k}^{2}+(1+d)^{2}m^{2}]\Psi_{\bf k}+\nonumber \\ \\
&& +4\left(\frac{g^{2}}{4\pi}\right)(\xi+1)|{\bf k}|\int d{\bf k}'
\frac{|{\bf k}'|}{[({\bf k}')^{2}+
(1+d)^{2}m^{2}]^{1/2}}\Psi_{{\bf k}'}=0\;\;.
\nonumber 
\end{eqnarray}

\vskip 1.0cm
\section{Two-body dynamics from the linearized 
mean-field equations} \vskip 0.7cm

In this section we implement and discuss a reinterpretation of the small
oscillations problem along the line proposed by Kerman and Lin in 
Ref.\cite{KL}. We begin rewriting equations (\ref{27}) and (\ref{28}) 
as

\begin{eqnarray}
\label{29}
&&\Gamma_{\bf k}=-\frac{i\omega}{md}
\frac{k_{0}}{|{\bf k}|}\Psi_{\bf k}
\\ \nonumber\\
\label{30}
&&\frac{(k_{0}^{\mbox{\scriptsize eff}})^{2}}
{k_{0}^{\mbox{\scriptsize eff}}}
\Psi_{\bf k}-\left(\frac{g^{2}}{4\pi}\right)(\xi+1)
\frac{|{\bf k}|}{(k_{0}^{\mbox{\scriptsize eff}})}
\int d{\bf k}'\frac{|{\bf k}'|}
{(k_{0}^{\mbox{\scriptsize eff}})'}
\Psi_{{\bf k}'}=\frac{\omega^{2}}{4k_{0}^{\mbox{\scriptsize eff}}}
\Psi_{\bf k} \;\;\;,
\end{eqnarray}
\vskip 0.5cm

\noindent where $(k_{0}^{\mbox{\scriptsize eff}})^{2}= [{\bf
k}^{2}+(1+d)^{2}m^{2}]=[{\bf k}^{2}+ {m_{\mbox{\scriptsize
eff}}^{2}}]$.

The crucial point is to realize that Eq.(\ref{30}) has the form of 
a Lippmann-Schwinger equation with separable potential

\begin{eqnarray}
\label{31}
&&\langle {\bf k}|V|{\bf k}'\rangle = v({\bf k})v^{*}({\bf k}') 
= \left(\frac{g^{2}}{4\pi}\right)(\xi+1)
h({\bf k})h({\bf k}')\nonumber\\ \\
&&{\rm where}\;\;\; h({\bf k})=
\frac{|{\bf k}|}{k_{0}^{\mbox{\scriptsize eff}}}\;\;.\nonumber
\end{eqnarray}
\vskip 0.5cm

In order to interpret this scattering problem we follow Ref.\cite{KL} in
relating dynamical small amplitude distortions of the Gaussian vacuum
to two-quasi-fermion states. The Gaussian vacuum $|\tilde 0\rangle$ can
be written explicitly in terms of the fermion operators $b^{\dag}
_{{\bf k},\lambda}$ and the Nambu parameters in the well-known Bardeen-Cooper-Schriffer (BCS) form

\begin{equation}
\label{31a}
|\tilde 0\rangle= \prod_{{\bf k}>0,\lambda\neq\bar\lambda}\left[
\cos \varphi_{\bf k}+(-1)^{\lambda}
\sin \varphi_{\bf k}e^{-i\gamma_{\bf k}}
b^{\dag}_{{\bf k},\lambda}b^{\dag}_{\bar{\bf k},\bar\lambda}\right]|0
\rangle\;\;,
\end{equation}
\vskip 0.5cm

\noindent
where $\varphi_{\bf k}=\varphi_{\bf k}|_{\rm eq}$ and $\gamma_{\bf k}=
\gamma_{\bf k}|_{\rm eq}$ and $|0\rangle$ is the vacuum of the $b$-operators. 
It is also well known that $|\tilde 0\rangle$ is the quasi-fermion  vacuum,
namely

\begin{equation}
\label{31b}
\beta_{{\bf k},\lambda}|\tilde 0\rangle=0
\end{equation} 

\noindent
and that one- and two-quasi-fermion states can be written as

\begin{equation}
\label{31c}
\beta^{\dag}_{{\bf k}_{1},\lambda_{1}}|\tilde 0\rangle=b^{\dag}_{{\bf 
k}_{1},\lambda_{1}}\prod_{({\bf k},\lambda)\neq({\bf k}_{1},\lambda_{1})}\left[
\cos \varphi_{\bf k}+(-1)^{\lambda}\sin \varphi_{\bf k}
e^{-i\gamma_{\bf k}}
b^{\dag}_{{\bf k},\lambda}b^{\dag}_{\bar{\bf k},\bar\lambda}\right]|0
\rangle
\end{equation}

\begin{eqnarray}
\label{31d}
\beta^{\dag}_{{\bf k}_{1},\lambda_{1}}\beta^{\dag}_{\bar{\bf k}_{1},
\bar\lambda_{1}}|\tilde 0\rangle&=&
\left[(-1)^{\bar\lambda_{1}}
\sin \varphi_{\bf k}e^{i\gamma_{\bf k}}+\cos \varphi_{\bf k}
b^{\dag}_{{\bf k}_{1},\lambda_{1}}b^{\dag}_{\bar{\bf k}_{1},\bar
\lambda_{1}}\right]\times\nonumber\\\nonumber\\
&\times &\prod_{({\bf k},\lambda)\neq({\bf k}_{1},\lambda_{1})}\left[
\cos \varphi_{\bf k}+(-1)^{\lambda}\sin \varphi_{\bf k}
e^{-i\gamma_{\bf k}}
b^{\dag}_{{\bf k},\lambda}b^{\dag}_{\bar{\bf k},\bar\lambda}\right]
|0\rangle
\end{eqnarray}
\vskip 0.5cm

On the other hand, making the variations Eq.(\ref{26}) in 
Eq.(\ref{31a}) leads to a first order variation of the Gaussian 
vacuum which, appart from the exponential time dependence, is given by 

\begin{eqnarray}
\label{31e}
\delta|\tilde 0\rangle&=&\Psi_{{\mbox{\scriptsize\bf k}}_{1}}
\beta^{\dag}_{{\bf k}_{1},\lambda_{1}}\beta^{\dag}_
{{\bar{\bf k}}_{1},{\bar
\lambda}_{1}}|\tilde 0\rangle - i (-1)^{\bar\lambda_{1}}
\Gamma_{{\mbox{\scriptsize\bf k}}_{1}}
\sin \varphi_{{\bf k}_{1}}|_{\rm eq}
b^{\dag}_{{\bf k}_{1},\lambda_{1}}b^{\dag}_{{\bar{\bf k}}_{1},{\bar
\lambda}_{1}}\times\nonumber\\\nonumber\\
&\times &\prod_{({\bf k},\lambda)\neq({\bf k}_{1},\lambda_{1})}
\left[
\cos \varphi_{\bf k}+(-1)^{\lambda}
\sin \varphi_{\bf k}e^{-i\gamma_{\bf k}}
b^{\dag}_{{\bf k},\lambda}b^{\dag}_{\bar{\bf k},\bar\lambda}\right]|0
\rangle\\ \nonumber\\\nonumber
\end{eqnarray}

\noindent
This result just illustrates the well-known theorem by Thouless \cite{RS}
when $\Gamma_{\mbox{\scriptsize\bf k}_{1}}=0$ (as is appropriate in 
a static
context), in which case it says that the linear variation corresponds to
two-paired quasi-fermion addition to the Gaussian vacuum. Kerman and Lin
have used this fact, in the context of the $\phi^4$ theory, to 
associate the
scattering problem (\ref{30}) to the two-boson dynamics. A 
similar association
can thus also be made here. Moreover, the last term of Eq.(\ref{31e})
indicates the presence of dynamical symmetry breaking effects, since this
term, proportional to the variation of the phase $\gamma_{\bf k}$, vanishes 
unless the chiral symmetry is broken in the static vacuum 
($\varphi_{\bf k}\neq 0$).

The small oscilation regime can thus be seen as a non-perturbative way
of approaching the dynamics of paired two-quasi-fermion excitations 
of the
vacuum, including dynamical symmetry breaking effects.

\subsection{Analytical solution of the linearized equations}

We will now show how Eq.(\ref{29}) and (\ref{30}) can be solved
analytically. A general solution to two-fermion wave function 
$\Psi_{\bf k}$ will
have two terms. The first one is the free solution ($g=0$ vanishing
potential) and represents an incident wave. The second term is the non
trivial part (when $g \neq 0$) which couples different momenta, and is
associated with the scattered wave. Thus

\begin{eqnarray}
\label{32}
&&\frac{|{\bf k}|}{k_{0}^{\mbox{\scriptsize eff}}}\Psi({\bf k},{\bf
q};\omega)= \alpha\delta({\bf q}-{\bf k}) + \nonumber \\ \\ &&+
\frac{1}{[({\bf k}_{0}^{\mbox{\scriptsize eff}})^{2}-\omega^{2}/4+
i\epsilon]}\left(\frac{g^{2}}{4\pi}\right)(\xi+1) \frac{{\bf
k}^{2}}{k_{0}^{\mbox{\scriptsize eff}}}\int d{\bf k}' \frac{|{\bf
k}'|}{(k_{0}^{\mbox{\scriptsize eff}})'} \Psi({\bf k}',{\bf
q};\omega)\;\;\;\;, \nonumber
\end{eqnarray}

\vskip 0.3cm
\noindent where ${\bf q}$ is interpreted as the 
relative momentum for two incident quasi-fermions and $\alpha$ is an
overall phase factor.  We choose the outgoing wave condition
$(+i\epsilon)$ as solution of Eq.(\ref{30}), but we could have chosen
e.g.  the incoming wave condition $(-i\epsilon)$ or Van Kampen wave
condition \cite{CPP} or another condition.
 
Integrating the Eq.(\ref{32}) with respect to ${\bf k}$ 

\begin{equation}
\label{33}
\int d{\bf k}\frac{|{\bf k}|}{(k_{0}^{\mbox{\scriptsize eff}})}
\Psi({\bf k},{\bf q};\omega)=\frac{\alpha}
{\left\{1-\left(g^{2}/4\pi\right)(\xi+1)\int \frac{d{\bf k}'}
{(k_{0}^{\mbox{\scriptsize eff}})'}\frac{({\bf k}')^{2}}
{[\{(k_{0}^{\mbox{\scriptsize eff}})'\}^{2}-
\omega^{2}/4+i\epsilon]}\right\}}
\end{equation}
\vskip 0.5cm

\noindent and substituting this result back into Eq.(\ref{32}) 
yields a general solution for $\Psi({\bf k},{\bf q};\omega)$ 

\begin{equation}
\label{34}
\frac{|{\bf k}|}{k_{0}^{\mbox{\scriptsize eff}}}\Psi({\bf k},{\bf
q};\omega)= \alpha\delta({\bf q}-{\bf k})+\frac{\alpha
k_{0}^{\mbox{\scriptsize eff}}} {[({\bf k}_{0}^{\mbox{\scriptsize
eff}})^{2}-\omega^{2}/4+i\epsilon]} \left(\frac{|{\bf
k}|}{k_{0}^{\mbox{\scriptsize eff}}}\right)\frac{1}
{\Delta^{+}(\omega)}\left(\frac{|{\bf k}|}{k_{0}^{\mbox{\scriptsize
eff}}} \right)\;\;,
\end{equation}
\vskip 0.5cm

\noindent where $\Delta^{+}(\omega)$ is given by 

\begin{equation}
\label{35}
\Delta^{+}(\omega)=\left(\frac{4\pi}{g^{2}}\right)\frac{1}{(\xi+1)}-
\int \frac{d{\bf k}}{(k_{0}^{\mbox{\scriptsize eff}})}
\frac{{\bf k}^{2}}{[(k_{0}^{\mbox{\scriptsize eff}})^{2}-
\omega^{2}/4+i\epsilon]}\;\;.
\end{equation}
\vskip 0.5cm

The oscillation amplitude $\Gamma({\bf k},{\bf q};\omega)$ is obtained
from the Eqs.(\ref{29}) and (\ref{34}) and reads

\begin{eqnarray}
\label{36}
&&\Gamma({\bf k},{\bf q};\omega)=
\left.-\frac{i\omega\alpha k_{0}}
{md}\left(\frac{k_{0}^{\mbox{\scriptsize eff}}}{k^{2}}\right)
\right.\left\{\delta({\bf q}-{\bf k})+
\right.\nonumber \nonumber\\ \\
&&+\left.\frac{k_{0}^{\mbox{\scriptsize eff}}}
{[(k_{0}^{\mbox{\scriptsize eff}})^{2}-
\omega^{2}/4+i\epsilon]}
\left(\frac{|{\bf k}|}{k_{0}^{\mbox{\scriptsize eff}}}\right)
\frac{1}{\Delta^{+}(\omega)}
\left(\frac{|{\bf k}|}{k_{0}^{\mbox{\scriptsize eff}}}
\right)\right\}\;\;. \nonumber 
\end{eqnarray}
\vskip 0.5cm

Finally, substituting Eq.(\ref{34}) into Eq.(\ref{30}) we obtain the
oscillation frequencies

\vskip 0.2cm
\begin{equation}
\label{37}
\omega=2q_{0}^{\mbox{\scriptsize eff}}=
2[q^{2}+m_{\mbox{\scriptsize eff}}^{2}]^{1/2}\;\;,
\end{equation}
\vskip 0.5cm

\noindent where ${\bf q}$ is the relative momentum 
for two incident quasi-fermion with masses $m_{\mbox{\scriptsize
eff}}=(1+d)m$.

We observe that we can understand the factor 2 in the frequencies of
oscillation $\omega$ [see Eq.(\ref{37})], as related to the treatment
of harmonic oscillators in terms of the sympletic groups given by
Goshen and Lipkin \cite{GL}.  It can be interpreted classically by
noticing that, since for harmonic oscillators the frequency does not
depend on the amplitude of the motion, if a set of independent
particles in a harmonic field is symmetrically stretched out of
equilibrium, it will subsequently pulsate with frequency $2\omega$,
where $\omega$ is the frequency of oscillation of the independent
particles.

\vskip 1.0cm
\section{Bound states from the small oscillations regime}
\vskip 0.7cm

In this section we will examine the condition for existence of bound
states in the small oscillation regime around the stationary solution
(vacuum) of our fermionic system.

From Eqs.(\ref{30}) and (\ref{31}) we verify that the potential term,
which describes the time evolution of our system in this regime is
separable. Again, in analogy with scattering theory, we can evaluate
the corresponding $T$ matrix \cite{RN}. We find

\begin{equation}
\label{38}
T({\bf k},{\bf k}';\omega)\propto h({\bf k})
\frac{1}{\Delta^{+}(\omega)}
h({\bf k}')
\end{equation}
\vskip 0.5cm

\noindent with $h({\bf k})$ given by Eq.(\ref{31}) and 
$\Delta^{+}(\omega)$ given by Eq.(\ref{35}).  The bound states are
given by the poles of the $T$ matrix.  Therefore, we search for the
zeros of $\Delta^{+}(\omega)$. It is clear that the integral in
$\Delta^{+}(\omega)$ contains a logarithmic divergence. To keep it
under control, we use the renormalization procedure of the coupling
constant given by Eq.(\ref{14}). Substituting Eq.(\ref{14}) into
Eq.(\ref{35}) we get

\begin{equation}
\label{39}
\Delta^{+}(\omega)=\ln\left(\frac{\Lambda^{2}}{m^{2}}\right) - 
\int_{-\Lambda}^{+\Lambda}\frac{d{\bf k}}{[{\bf k}^{2}+
m_{\mbox{\scriptsize eff}}^{2}]^{1/2}}
\frac{{\bf k}^{2}}{[{\bf k}^{2}+
m_{\mbox{\scriptsize eff}}^{2}-\omega^{2}/4+i\epsilon]}\;\;\;,
\end{equation}
\vskip 0.5cm

\noindent where we introduce the regularizing momentum cut-off
$\Lambda$.

In the interval $0\leq \omega\leq 2m_{\mbox{\scriptsize eff}}$ 
the integral of $\Delta^{+}(\omega)$ is well defined 
and we can set $\epsilon=0$. A straightforward calculation yields

\begin{equation}
\label{40}
\Delta^{+}(\omega)=2\left[f(\omega)-j(d)\right]
\end{equation}

\begin{eqnarray}
\label{41}
{\rm where}\;\;\;f(\omega)
&=&\left[\frac{4m_{\mbox{\scriptsize eff}}^{2}}
{\omega^{2}}-1\right]^{1/2}
\arctan\left\{\left[\frac{4m_{\mbox{\scriptsize eff}}^{2}}
{\omega^{2}}-1\right]^{-1/2}\right\}\nonumber\\ \\
j(d)&=& \ln\left[\frac{2}{|1+d|}\right]\;\;.\nonumber\\ \nonumber
\end{eqnarray}

Fig.1 shows the zero of the $\Delta^{+}(\omega)$ as function of
$d$. In this calculation ${\bf q}=0$, therefore $\omega$ is the mass
of the bound state. Obviously, when $(1+d)=0$ or $m_{\mbox{\scriptsize
eff}}=0$ (free system, see Ref.\cite{NTP}) there is no bound state.
We see from Fig.1 that a bound state of quasi-fermions occurs when
$0.74\leq(1+d)\leq2$, and that the mass of this bound state will vary
in the interval $0\leq\omega\leq 2m_{\mbox{\scriptsize eff}}$.  Gross
and Neveu obtain $M_{\sigma}=2M_{F}$ \cite{GN} for the mass of the
$\sigma$ particle in leading-$1/N$ aproximation, where $M_{F}$ is
equivalent to $m_{\mbox{\scriptsize eff}}$. They argue that in higher
order they might find that

\[
M_{\sigma}=2M_{F}\left[1+{\cal O}\left(1/N\right)\right]\;\;.
\]

\noindent From Fig.1 we verify that $\omega=2m_{\mbox{\scriptsize
eff}}$ corresponds to $(1+d)=2$.  Observing that $j(1+d=2)=0$, we may
conclude that $j(d)$ can be see as a contribuition of higher order to
the Gross-Neveu result.

We believe that in the limit $N\rightarrow\infty$ the function
$j(d)\rightarrow 0$. On the other hand, when $N$ is finite [$N=1$ in
our calculation - see Eq.(\ref{10})], the mass $\omega$ of the bound
state depends of the renormalized coupling constant $d$ as shown in
Fig.1.  This dependence can not be obtained from $1/N$ approximation.

Therefore, we can conclure that to $N\rightarrow\infty$ the $1/N$
approximation and our mean-field approximation are equivalent. On the
other hand, when $N$ is finite our approximation permits to obtain the
higher order contribution to the Gross-Neveu result \cite{GN}.

It is important to observe that the higher order term obtained in
Eqs.(\ref{40}) and (\ref{41}) in this approach does not contain all
terms of order $1/N$, since the mean-field approximation is not a
systematic expansion in the parameter $1/N$.

Surprisingly, we have obtained for the function $\Delta^{+}(\omega)$ a
structure which entirely reproduces that which has been found by
Kerman and Lin \cite{KL} in their study of the {\it bosonic}
$\lambda\phi^{4}$ theory in terms of a Gaussian time-dependent
variational approach.

Finally, when $\omega>2m_{\mbox{\scriptsize eff}}$ the 
integrand of $\Delta^{+}(\omega)$ has a singularity at 
${\bf k}=\pm\sqrt{\omega^{2}/4-m_{\mbox{\scriptsize eff}}^{2}}$ . 
From the theory of residues we obtain 

\begin{eqnarray}
\label{42}
\Delta^{+}(\omega)&=&\left(1-\frac{4m_{\mbox{\scriptsize eff}}^{2}}
{\omega^{2}}\right)^{1/2}
\ln\left[\frac{1+\left(1-4m_{\mbox{\scriptsize eff}}^{2}/
\omega^{2}\right)^{1/2}}{1-
\left(1-4m_{\mbox{\scriptsize eff}}^{2}/
\omega^{2}\right)^{1/2}}\right]+\nonumber \\ \\
&-&2\ln\left(\frac{2}{|1+d|}\right)
-i\pi\left[1-\frac{4m_{\mbox{\scriptsize eff}}^{2}}
{\omega^{2}}\right]^{1/2}\;\;.\nonumber\\ \nonumber
\end{eqnarray}

Now $\Delta^{+}(\omega)$ does not have any zeros.  The interesting
point here is to observe that

\begin{equation}
\label{43}
\lim_{\omega\rightarrow\infty}\Delta^{+}(\omega)\longrightarrow 
\ln\left(\frac{\omega^{2}}{m_{\mbox{\scriptsize eff}}^{2}}\right)
\longrightarrow\infty \;\;,
\end{equation}
\vskip 0.5cm

\noindent so that in the large frequency limit 
$(\omega\rightarrow\infty)$ the $T$ matrix goes asymptotically to
zero. We thus recover, in the present approximation, the
assymptotically free character of the CGNM.

\vskip 1.0cm 
\section{Discussion and conclusions}
\vskip 0.7cm

In Ref.\cite{NTP} we described a treatment of the initial-values
problem in a quantum field theory of self-interacting fermions in the
Gaussian approximation. Although the procedure is quite general, we
implemented it for the vacuum of an uniform (1+1) dimensional
relativistic many-fermion system described by Chiral Gross-Neveu model
(CGNM). We obtained the renormalized kinetic equations which describe
the effective dynamics of the Gaussian observables in the mean-field
approximation for this system.

In this work, we have considered the linearized form of the mean-field
kinetic equations obtained in Ref.\cite{NTP} around the stationary
(vacuum) solution. The two-quasi-fermion physics can be analytically
investigated in this approach. In particular, we have solved these
equations completely.  From the solutions, we have reinterpreted the
near equilibrium physics of our system as a problem of quasi-fermion
scattering and have found the condition for the existence of a
quasi-fermion bound state.

We verify that for $N$ finite (in this work $N=1$), the bound state
mass obtained from our approach contains a term which depends of the
renormalized coupling constant as can be seen in the Fig.1. In the
case of an $1/N$ expansion \cite {GN} this dependence can not be
found, so in the limit $N\rightarrow\infty$ this term goes to
zero. Therefore, to small $N$, our approach permits to obtain the
higher order contribuition to the $1/N$ expansion.

Finally, it is important to observe that the higher order term
obtained to the bound state mass from our approach contains no
necessarily all terms of $1/N$ order, since the mean-field
approximation is not an $1/N$ expansion.

\vskip 1.0cm
\centerline{\Large\bf Acknowledgments}
\vskip 0.7cm

One of the authors (P.L.N.) was 
supported by Conselho Nacional de 
Desenvolvimento Cient{\'{\i}}fico e 
Tecnol\'ogico (CNPq), Brazil; and by 
Funda{\c{c}\~ao} de Amparo \`a Pesquisa 
do Estado de S\~ao Paulo (FAPESP), 
Brazil.

\vskip 0.7cm

\centerline{\Large\bf Figure Captions}
\vskip 0.7cm

Figure 1 - The curve represents the mass $\omega$ of the 
two-quasi-fermion bound state in small oscillation regime 
for our system as a function of the renormalized coupling 
constant $d$.

\end {document}